\pdfoutput=1
\documentclass[10pt,twocolumn,letterpaper]{article}

\usepackage[pagenumbers]{cvpr}
\makeatletter\patchcmd{\@maketitle}{\vskip .375in}{\vskip .1in}{}{\typeout{WARNING: title-space patch failed}}\makeatother

\usepackage{amsmath,amssymb}
\usepackage{booktabs}
\usepackage{subcaption}
\usepackage{graphicx}

\newcommand{\red}[1]{#1}

\usepackage{multirow}
\usepackage{colortbl}

\usepackage{balance}

\definecolor{cvprblue}{rgb}{0.21,0.49,0.74}
\usepackage[pagebackref,breaklinks,colorlinks,allcolors=cvprblue]{hyperref}

\def\paperID{*****}
\def\confName{CVPR}
\def\confYear{2027}

\title{LiveFoley: Dynamic Real-Time Video-to-Audio Generation\\for Immersive Interactive Worlds}

\author{
Yusheng Dai$^{1,3}$ \quad Vladimir Iashin$^{4}$ \quad Kangdi Wang$^{5}$ \quad Yuxuan Jiang$^{6}$ \quad Ilpo Viertola$^{4}$\\
Weiqiang Wang$^{1}$ \quad Qiuhong Ke$^{1}$ \quad Min Zhao$^{2,\dagger}$ \quad Jianfei Cai$^{1}$\\[2pt]
{\normalsize $^{1}$Monash University \quad $^{2}$Nanjing University \quad $^{3}$ShengShu Technology \quad $^{4}$Tampere University}\\
{\normalsize $^{5}$Alibaba Token Hub, Alibaba Group \quad $^{6}$Tsinghua University \quad {\small $^{\dagger}$Corresponding author}}
}

\begin{document}
\maketitle
\begin{abstract}
Streaming video-to-audio (V2A) generation is essential for immersive interactive applications. However, real-time constraints leave existing streaming V2A methods far behind offline models: (1) visual frontends shrink their receptive fields to cut look-ahead latency, severely degrading audio-visual alignment; (2) \red{causal autoregressive diffusion backends denoise each audio patch in isolation}, impairing audio quality.
To this end, we present LiveFoley, a streaming causal autoregressive diffusion architecture that addresses both bottlenecks with imperceptible latency. For the visual frontend, we introduce a Trailing Sliding Window that converts non-causal segment-level encoders into streaming modules without retraining. By simply realigning the window, it \red{preserves} precise audio-visual alignment\red{, while a configurable hop size reduces the frontend look-ahead well below the human perceptual threshold}. Building on this tunable hop size, we propose the Unified Varhop Framework to accommodate scenarios with diverse latency and computation budgets. For the audio backend, \red{we explore widening the denoising window of the \red{local diffusion transformer (LocDiT)} head in two temporal directions. History LocDiT adds past context to improve quality at no latency cost and serves as our default. Alternatively, Future-Warmup LocDiT breaks the autoregressive deadlock with an approximate future condition, further improving generation quality at the cost of an acceptable backend look-ahead.} Extensive experiments demonstrate that LiveFoley matches offline models in \red{temporal} alignment \red{and} leads streaming systems in generation quality. Project page: \url{https://dalision.github.io/livefoley-page/}.
\end{abstract}

\section{Introduction}
\label{sec:intro}

In interactive virtual environments, real-time acoustic feedback to user actions is crucial for immersion. While recent world models~\cite{gwmworlds2,helixworld,echowm} have advanced synchronized audio-visual generation, their audio synthesis is inherently coupled with their internal video generation, rendering them incompatible with external visual pipelines such as game engines, \red{AR/VR}, and live camera feeds. In contrast, streaming video-to-audio (V2A) generation presents a versatile, source-agnostic alternative by synthesizing audio on-the-fly from arbitrary live pixel streams. Unlike turn-based systems (\eg, GPT-Realtime~\cite{gptrealtime}), streaming V2A requires continuous and precise temporal synchronization\red{, so visual encoding and audio synthesis must jointly execute within an ultra-low latency budget to keep the audio-visual delay below the human perceptual threshold.}

\begin{figure}[t]
  \centering
  \includegraphics[width=\linewidth]{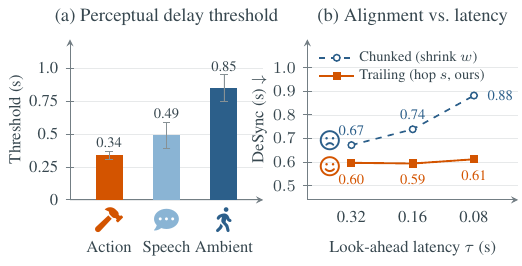}
  \caption{(a)~User study: the minimum audio delay at which participants notice the sound lagging behind the video, for action sounds, speech, and ambient sounds in free roaming (mean $\pm$ 95\% CI). (b)~Empirical study of streaming visual frontends on the V-AURA~\cite{vaura} backbone: the traditional chunked window (Figure~\ref{fig:visual_frontend}a) cuts latency by shrinking the window $w$ at a sharp cost in audio-visual alignment, while our trailing sliding window (Figure~\ref{fig:visual_frontend}b) shortens the hop $s$ and preserves alignment.}
  \vspace{-15pt}
  \label{fig:motivation}
\end{figure}

To quantify this \red{perceptual} threshold, \red{we conduct a} user study (Figure~\ref{fig:motivation}a; protocol in Appendix~\ref{sec:user_study})\red{, which} reveals two key observations. First, discrete action-sound events (\eg, impacts) are the most sensitive: delays become perceptible at about 0.34\,s\red{\footnote{In Appendix~\ref{sec:user_study}, we compare the difference from ITU-R BT.1359~\cite{itu_bt1359} that our reference-free protocol better reflects real interactive use.}}, which sets a strict latency budget for preserving immersion. Second, the threshold varies sharply with context: delays in speech are perceived at about 0.49\,s, whereas ambient sounds during free roaming are only perceived at about 0.85\,s. These findings \red{suggest} that an ideal streaming V2A system must (1) achieve high audio-visual alignment and generation quality within an ultra-low latency budget, and (2) dynamically adapt to different scenarios, trading allowable delay for reduced computational load.

However, existing streaming V2A approaches cannot achieve imperceptible latency, precise audio-visual alignment, and high audio quality at the same time, owing to two core bottlenecks. \textit{The first bottleneck is the look-ahead delay of the visual frontend.} Precise audio-visual alignment typically requires segment-level encoders (\eg, Synchformer~\cite{synchformer}), but their full segment encoding introduces a long look-ahead delay (\eg, 0.64\,s in V-AURA~\cite{vaura}). Conversely, frame-level encoders (\eg, DINOv2~\cite{dinov2}), as adopted in SoundReactor~\cite{soundreactor}, \red{require only a 0.08\,s frontend look-ahead} but severely \red{degrade audio-visual} alignment. Our controlled experiment (Figure~\ref{fig:motivation}b) further reveals the root cause: as the receptive field of Synchformer shrinks, the audio-visual alignment of downstream generation degrades steadily, showing that alignment hinges on the temporal context each segment covers. \textit{The second bottleneck stems from the causal constraints of continuous autoregressive generation.} Unlike offline diffusion models that jointly denoise all patches, \red{existing streaming V2A systems~\cite{soundreactor,worldsonus}, built on autoregressive diffusion architectures~\cite{mar,ditar,voxcpm}, denoise each patch in isolation with a local diffusion or flow-matching head (LocDiT), attending to neither previous nor upcoming patches. As a result,} these streaming models significantly trail offline systems in audio quality.

To address these challenges, we propose \textit{LiveFoley}, a plug-and-play sound engine for general vision-interactive applications that delivers real-time, precisely synchronized, and high-quality acoustic feedback. For the visual frontend, we introduce a simple yet effective mechanism, the \textbf{Trailing Sliding Window}, that breaks the latency-alignment trade-off. Rather than shrinking the receptive field, we anchor the tail of the segment window to the current frame. Paired with first-frame padding, this mechanism converts off-the-shelf non-causal segment-level encoders into streaming modules, decoupling the \red{frontend look-ahead} from the long window length and instead linking it to a small, configurable hop size. Building on this tunable hop size, we further propose the \textbf{Unified Varhop Framework} to accommodate the diverse latency and compute budgets of different scenarios (\eg, free roaming vs. shooter games). At its core, a Synchronizer module aligns the latent representations across different latency modes, enabling a shared generation backbone to adopt dynamic hop sizes.

For the audio backend, \red{we explore widening the denoising window of LocDiT beyond a single patch. Toward the past, \textbf{History LocDiT} adds previously generated patches as context, improving quality at no latency cost, and serves as our default. Toward the future, the upcoming patch cannot simply join the current denoising window, as its condition depends on the clean current patch, which is not yet available. \textbf{Future-Warmup LocDiT} breaks this autoregressive deadlock by constructing an approximate condition from the current hidden state and the next visual token, enabling cross-patch collaborative denoising. As an optional upgrade, it improves quality further with only 0.08\,s of extra backend look-ahead, keeping the \red{end-to-end latency at an estimated} 236\,ms (\red{with hop size} $s=1$, Appendix~\ref{sec:latency_appendix}), below the human perceptual threshold (Figure~\ref{fig:motivation}a).}

Experiments show that \textit{LiveFoley} is the first streaming V2A system to match offline models in \red{temporal} alignment, with only a 0.08\,s \red{frontend look-ahead}, while delivering the best audio quality among streaming systems. On a single A800 GPU, it achieves an \red{end-to-end latency} below the human perceptual threshold. \textit{LiveFoley} can thus serve as a plug-and-play sound engine for interactive applications such as AR/VR and video games.

\section{Related Work}
\label{sec:related}

\begin{figure*}[t]
  \centering
  \includegraphics[width=\textwidth]{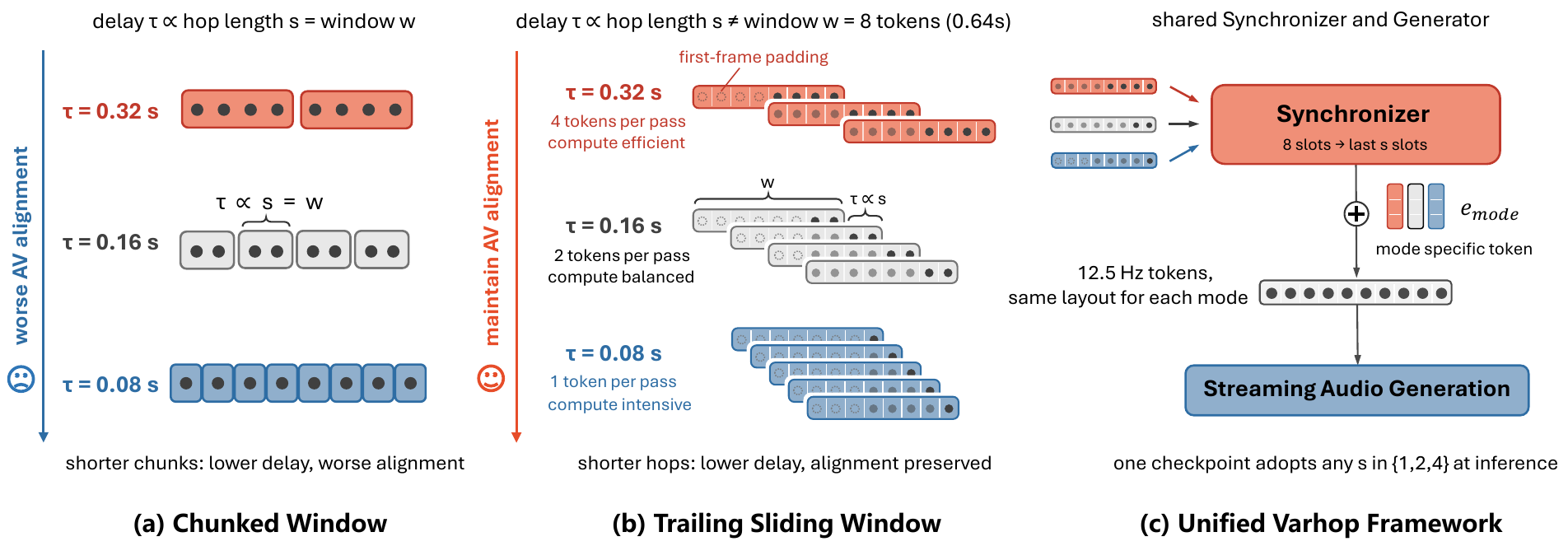}
  \caption{Streaming visual frontend. Each dot denotes the token at one window position. (a)~Chunked window: shrinking the window $w$ lowers latency but also narrows the receptive field, which degrades audio-visual alignment. (b)~Trailing sliding window: anchoring the window tail to the latest frame reduces look-ahead to one hop while preserving the full receptive field $w$ and hence the alignment. Black dots are emitted tokens at newly arrived frames. Gray and dashed dots correspond to earlier frames and first-frame padding, respectively; these frames provide context only within the encoder, and their tokens are discarded. (c)~Unified Varhop Framework: a Synchronizer maps tokens from different hop sizes into a shared latent space, allowing a single backend to switch among hop sizes.}
  \label{fig:visual_frontend}
  \vspace{-10pt}
\end{figure*}

\paragraph{Audio-Visual Alignment for Streaming V2A.}
In offline video-to-audio (V2A) generation, visual representations have evolved from coarse per-frame features, such as CLIP~\cite{radford2021clip,im2wav}, RGB, and optical flow~\cite{specvqgan}, to segment-level features like Synchformer~\cite{synchformer}. By substantially improving temporal alignment, this advancement underpins leading offline systems (\eg, MMAudio~\cite{mmaudio}, HunyuanVideo-Foley~\cite{hunyuanfoley}). In streaming, however, these non-causal encoders introduce a latency-alignment trade-off: V-AURA~\cite{vaura} adopts Synchformer at the cost of a 0.64\,s look-ahead, whereas SoundReactor~\cite{soundreactor} uses per-frame DINOv2~\cite{dinov2} features with \red{minimal} look-ahead but weaker alignment. Concurrently, WorldSonus~\cite{worldsonus} partially mitigates this trade-off by aligning its autoregressive backbone to a frozen Synchformer teacher via a time-shifted contrastive loss, improving alignment under chunk-level (100\,ms) latency but still trailing offline models (Section~\ref{sec:main_results}).

\paragraph{Generative Architectures for Streaming V2A.}
While global bidirectional diffusion models dominate offline generation~\cite{mmaudio,thinksound,hunyuanfoley}, streaming applications necessitate autoregressive (AR) architectures to bypass the dependency on full-sequence video inputs. Early AR audio models rely on discrete codec tokens, predicting RVQ token streams autoregressively~\cite{audiogen,musicgen,foleygen,vaura}. However, discrete tokens suffer from quantization loss and require complex scheduling to manage intra-frame dependencies~\cite{mar,musicgen}. Recent approaches have therefore shifted toward continuous AR generation, pairing a causal LM with a local flow-matching or diffusion head~\cite{mar,ditar,voxcpm,dotstts,rouard2025calm,soundreactor,swansphere,worldsonus}. By eliminating quantization bottlenecks, this continuous paradigm \red{can capture} finer acoustic details. \red{Yet streaming V2A models built on this paradigm still fall short of offline models in generation quality~\cite{soundreactor,worldsonus}. We attribute this partly to their local heads, which denoise each patch in isolation, see history only through the AR hidden state, and cannot denoise jointly with upcoming patches due to causality.}

\paragraph{Other Online Audio Generation.}
Beyond streaming V2A, the broader landscape of real-time interactive audio encompasses joint audio-visual generation and acoustic scene modeling. Joint generation models~\cite{gwmworlds2,helixworld,echowm,omniforcing,avwm} synchronously produce video and audio. However, their audio generation remains strictly coupled to internal video generators, rendering them inapplicable to pre-existing visual streams like game engines or camera feeds. Alternatively, acoustic scene modeling systems~\cite{sonoworld} focus purely on spatial rendering. These systems adjust the spatial attenuation of static environmental sounds based on user pose, yet cannot generate dynamic action sounds corresponding to transient visual events. In contrast, streaming V2A offers a plug-and-play solution that generates synchronized sound directly conditioned on arbitrary dynamic pixel streams, providing maximum flexibility for interactive scenarios.

\section{Method}
\label{sec:method}

\subsection{Preliminaries: Synchformer}
\label{sec:prelim}
Synchformer~\cite{synchformer} is a self-supervised audio-visual synchronization model used as the temporal condition in V2A systems~\cite{mmaudio,vaura}. It splits a video into segments of 16 frames (0.64\,s at 25\,fps) and encodes each segment with a visual encoder (Motionformer~\cite{motionformer}) and an audio encoder (AST~\cite{ast}), yielding 8 visual tokens per segment, \ie, one every 0.08\,s. The encoders are first pretrained with a segment-level contrastive objective (Segment AVCLIP),
\begin{equation}
\mathcal{L}_{\mathrm{AVCLIP}} = -\frac{1}{B}\sum_{i=1}^{B}\log\frac{\exp(\tilde{a}_i\cdot\tilde{v}_i/\kappa)}{\sum_{j=1}^{B}\exp(\tilde{a}_i\cdot\tilde{v}_j/\kappa)},
\label{eq:avclip}
\end{equation}
where $\tilde{a}_i$ and $\tilde{v}_i$ are the audio and visual features of segment $i$, $B$ is the batch size, and $\kappa$ is a learnable temperature\red{; the symmetric visual-to-audio term is omitted for brevity}. A lightweight transformer is then trained on the concatenated segment features to classify the audio-visual offset with the encoders frozen. \textit{LiveFoley} adopts the frozen Segment AVCLIP visual encoder of Synchformer as its frontend. Since attention within a segment is bidirectional, a token becomes available only after its entire 16-frame segment has arrived; removing this look-ahead without shrinking the receptive field is the goal of Section~\ref{sec:visual_frontend}.

\subsection{Streaming Visual Frontend}
\label{sec:visual_frontend}

\paragraph{Chunked Window.}
Existing streaming V2A methods divide the video into fixed, non-overlapping windows and feed them sequentially to the visual encoder (Figure~\ref{fig:visual_frontend}a). For Synchformer, a window of $w$ tokens spans $2w$ frames, all of which must be buffered before encoding, so the \red{frontend look-ahead} grows linearly with $w$. Shrinking the window from 4 tokens to 1 cuts \red{it} from 0.32\,s to 0.08\,s, but also narrows the temporal receptive field, and downstream audio-visual alignment drops sharply (Figure~\ref{fig:motivation}b). The chunked window thus trades alignment for latency.

\paragraph{Trailing Sliding Window.}
The core challenge therefore lies in reducing latency without sacrificing the receptive field $w$. To this end, \textit{LiveFoley} introduces a \textit{trailing sliding window} (Figure~\ref{fig:visual_frontend}b). Instead of shrinking $w$, we anchor the tail of the window to the current frame $f_n$, formulating the input sequence as $\mathcal{F}_n = [f_{n-2w+1}, \dots, f_n]$. By restricting the context to strictly historical frames, this design \red{avoids waiting} for a full window, decoupling the \red{frontend look-ahead} from $w$. Furthermore, when $n < 2w$, we pad the missing history by repeating the first frame, so that the first output does not have to wait for a full window of $2w$ real frames to accumulate, thereby reducing the \red{end-to-end latency}. Consequently, this formulation allows off-the-shelf non-causal segment-level encoders to operate causally in streaming settings without retraining, while preserving their bidirectional temporal attention.

In continuous streaming, the trailing window slides forward as new frames arrive. We introduce a hop size $s$ that governs the sliding frequency: as illustrated in Figure~\ref{fig:visual_frontend}b with Synchformer, the window advances every $2s$ frames, emitting exactly $s$ newly extracted trailing tokens. By adjusting $s$, we can precisely control the \red{frontend look-ahead} $\tau$ (\eg, $s \in \{1, 2, 4\}$ yields $\tau \in \{0.08, 0.16, 0.32\}$\,s). Unlike shrinking \red{the} window size $w$, reducing \red{the} hop size $s$ keeps the full temporal context, so latency drops while audio-visual alignment is preserved (\red{orange line in Figure~\ref{fig:motivation}b}). The latency-alignment dilemma thus becomes a configurable latency-compute trade-off.

\subsection{Unified Varhop Framework}
\label{sec:varhop}
As shown in Figure~\ref{fig:motivation}a, the tolerable audio delay varies widely across scenarios. Scenarios with strict audio-visual coupling (\eg, shooting or rhythm games) require a minimal \red{latency} ($\tau$) and a small hop size ($s$), whereas in ambient scenes (\eg, open-world roaming) delays are only perceived at about 0.85\,s, so a larger $s$ reduces the frontend extraction frequency and saves computation without compromising immersion.

Instead of training a separate expert model for each hop size, \textit{LiveFoley} introduces the \textit{Unified Varhop Framework} (Figure~\ref{fig:visual_frontend}c), which supports multiple latency modes with a unified model. However, extracting different numbers of trailing tokens from a fixed-size window shifts the feature distribution across hop sizes, which can disrupt the downstream audio backend. To map these tokens into a unified hidden space, we introduce a \textit{Synchronizer} module to bridge the frontend and backend. Given the 8 tokens $W \in \mathbb{R}^{8 \times d}$ of a Synchformer window, the \textit{Synchronizer} produces:
\begin{equation}
V_s = \operatorname{Synchronizer}(W)[-s:] + e_{\mathrm{mode}},
\label{eq:synchronizer}
\end{equation}
where $\operatorname{Synchronizer}(\cdot)$ is a two-layer transformer. Rather than directly \red{slicing the trailing} tokens from the raw frontend output, this module first incorporates positional embeddings and aggregates intra-window context for global feature alignment. Only then does it slice the final $s$ tokens and add a learnable latency-mode embedding, $e_{\mathrm{mode}}$. The output is a constant-rate 12.5\,Hz token stream for every hop size, so the audio backend is mode-agnostic and the hop size can be switched per scene at inference. During training, the three hop sizes are rotated in equal proportion across batches, so that a single set of weights learns all latency modes.

\subsection{Audio Generation Backbone}
\label{sec:backbone}
\begin{figure*}[t]
  \centering
  \includegraphics[width=\textwidth]{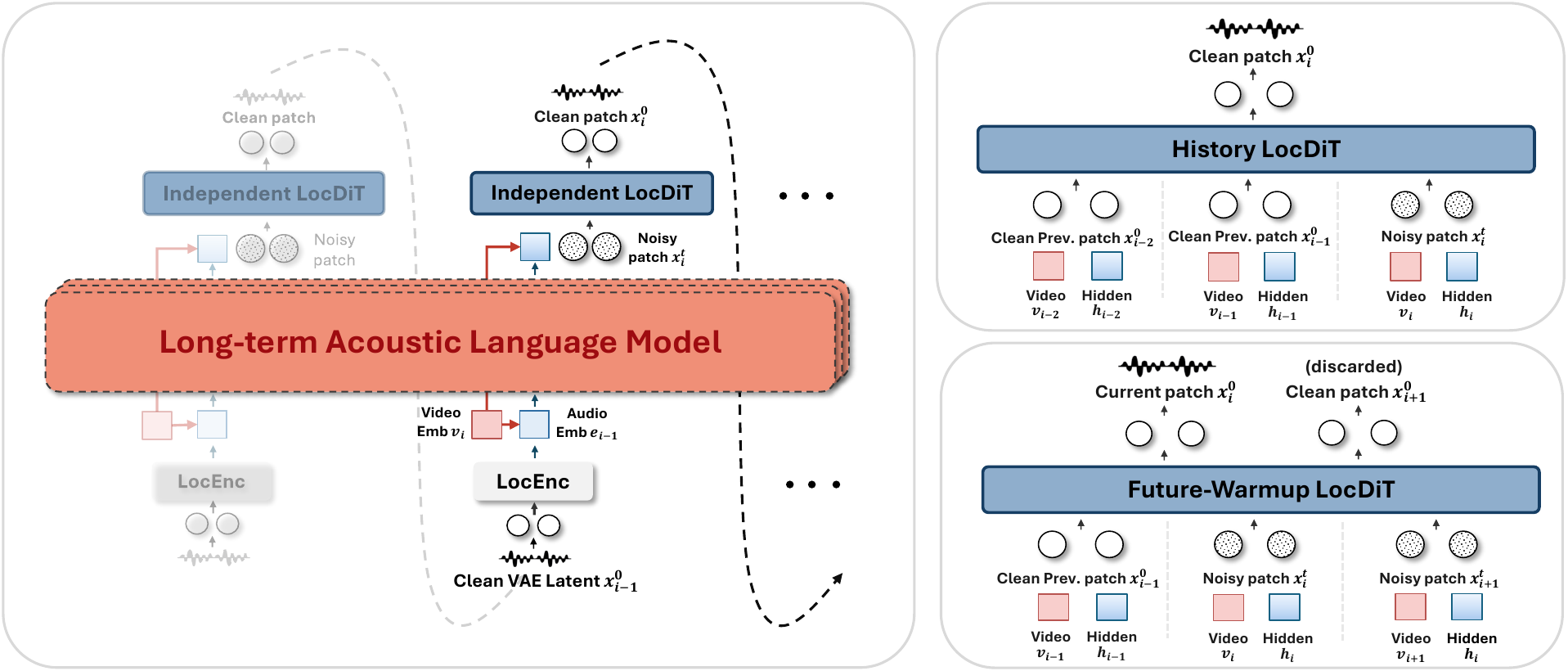}
  \caption{Continuous autoregressive audio generation backbone. Left: at each 12.5\,Hz step, the causal LM fuses the current visual token \red{$v_i$} and the LocEnc embedding \red{$e_{i-1}$} of the previous clean patch \red{$x_{i-1}^0$} into a hidden state $h_i$, on which \red{Independent} LocDiT denoises the current patch \red{$x_i^t$} \red{in isolation. Right: History LocDiT (top) adds clean patches \red{$x_{i-2}^0$ and $x_{i-1}^0$} as context without extra look-ahead. Future-Warmup LocDiT (bottom) jointly denoises \red{$x_i^t$} and a warmup patch \red{$x_{i+1}^t$}, which is conditioned on $h_i$ and $v_{i+1}$ and discarded at inference.}}
  \label{fig:audio_backbone}
\end{figure*}

\textit{LiveFoley} adopts an end-to-end \red{autoregressive diffusion} architecture operating on continuous representations from a causal audio VAE. As shown in Figure~\ref{fig:audio_backbone}, the generation module consists of two components: a causal language model (LM) and a local flow-matching transformer (LocDiT). Operating at a 12.5\,Hz autoregressive rate\red{, \ie, one step per visual sync token (0.08\,s)}, the acoustic LM captures long-term inter-patch dependencies to produce a coarse hidden state $h_i$. Conditioned on $h_i$, LocDiT \red{uses} bidirectional attention to refine intra-patch acoustic details. It acts as a local denoiser, transforming \red{a} noise patch $x_i^{1} = \epsilon$ into a clean VAE latent patch $x_i^{0}$, which is fed back to the LM for the next autoregressive step. Superscripts denote flow time, and $x_i$ without one refers to patch $i$. \red{Our framework supports both 16\,kHz and 48\,kHz causal VAEs, which share the same 12.5\,Hz patch rate: the 16\,kHz VAE outputs latent frames at 25\,Hz and groups $P=2$ frames into one patch, while the 48\,kHz VAE outputs 50\,Hz latents with $P=4$. We describe the 16\,kHz variant for simplicity.}

\paragraph{Causal Acoustic Language Model.}
Following \red{VoxCPM2}~\cite{voxcpm,voxcpm2}, we adopt the MiniCPM~\cite{minicpm,minicpm4} \red{architecture} as the LM backbone and \red{train it from scratch}. To stabilize the continuous autoregressive rollout, we employ a LocEnc module\red{~\cite{ditar,voxcpm}} that compresses the $P$ latent frames of the previously generated patch ($x_{i-1}^{0}$) into a single audio embedding $e_{i-1}$. Since the sync tokens from our visual frontend are already at 12.5\,Hz, they are injected without temporal resampling. At step $i$, the current sync token $v_i$ and the \red{previous} audio embedding $e_{i-1}$ are linearly projected \red{and} concatenated along the channel dimension. \red{The fused representation is then passed through a linear layer and fed into the LM to yield \red{the hidden state} $h_i$}.

\paragraph{Local Flow-Matching Head.}
LocDiT is a lightweight diffusion transformer responsible for generating the clean audio VAE patch at the current step. \red{For each denoised VAE token, we construct a condition vector $c_i$} by concatenating the LM hidden state $h_i$, the current visual sync token $v_i$, and the timestep embedding $\phi(t)$:
\begin{equation}
c_i = \mathrm{Concat}(h_i,\, v_i,\, \phi(t)).
\label{eq:locdit_cond}
\end{equation}
This condition is injected into the LocDiT blocks via adaptive layer normalization (adaLN)~\cite{dit}. \red{In prior streaming V2A work~\cite{soundreactor,worldsonus}, LocDiT denoises each patch in isolation and sees history only through $h_i$\red{;} we thus refer to this design as \textit{Independent LocDiT} (Figure~\ref{fig:audio_backbone}, left). Without cross-patch interaction, its bidirectional attention is confined to a single patch and lacks sufficient context, which \red{leads} to \red{less stable} training and poor acoustic quality.}

\subsection{\texorpdfstring{\red{Widening the Denoising Window}}{Widening the Denoising Window}}
\label{sec:window}
\red{Independent LocDiT and global bidirectional DiT (offline) mark the two extremes of cross-patch interaction. Between these extremes, we explore widening the denoising window of LocDiT toward the past and the future (Figure~\ref{fig:audio_backbone}, right).}

\paragraph{\red{Toward the Past: History LocDiT.}}
\red{History LocDiT widens the window unidirectionally toward the past by prepending the previously generated clean patches $x_{i-2}^{0}$ and $x_{i-1}^{0}$ to the LocDiT input as context. This gives its bidirectional attention more context and improves generation quality, without introducing extra backend look-ahead. The interaction, however, still flows only from past to present.}

\paragraph{\red{Toward the Future: Future-Warmup LocDiT.}}
\red{Adding the upcoming patch $x_{i+1}$ to the window makes the interaction between the current and upcoming patches bidirectional. This inevitably requires $v_{i+1}$ and thus introduces a 0.08\,s backend look-ahead, which is affordable thanks to the trailing sliding window. However, the fundamental obstacle is an autoregressive deadlock: $x_{i+1}$ is conditioned on $h_{i+1}$, which depends on $\mathrm{LocEnc}(x_i^{0})$ and thus cannot be computed while $x_i$ is still being denoised in the same window. A direct workaround, \textit{Low-Rate LocDiT}, halves the autoregressive rate so that each step denoises two patches jointly, but it also halves the temporal resolution of the backbone and degrades alignment (Section~\ref{sec:ablation}).}

\red{To break the deadlock without lowering the autoregressive rate of the LM, we construct \red{an approximate condition} for the future patch by pairing the current LM hidden state $h_i$ with \red{the} future visual token $v_{i+1}$:}

\begin{equation}
\begin{aligned}
c_i &= \mathrm{Concat}(h_i,\, v_i,\, \phi(t)),\\
c_{i+1} &= \mathrm{Concat}(h_i,\, v_{i+1},\, \phi(t)).
\end{aligned}
\label{eq:future_warmup}
\end{equation}

\red{This approximation resolves the circular dependency, so the upcoming patch $x_{i+1}$ can join the current window for joint denoising. Within this window, the two patches interact bidirectionally rather than following a strictly past-to-future flow: the current patch aligns with the acoustic trajectory of the future patch, while the future patch benefits from the context already established by the current one. During training, both $x_{i}$ and $x_{i+1}$ contribute to the loss. At inference, $x_{i+1}$ serves only as a warmup: it is discarded and regenerated at the next step with the true $h_{i+1}$.}

\begin{table*}[t]
\centering
\footnotesize
\setlength{\tabcolsep}{1.5mm}
\setlength{\aboverulesep}{1pt}
\setlength{\belowrulesep}{1pt}
\caption{System-level comparison on \red{the} VGGSound test set. \red{Frontend look-ahead is the duration of video that must be buffered before the visual encoder can encode a new chunk or hop}; offline models see the entire clip before generating. MMAudio (online) is our training-free online adaptation of MMAudio-L (Appendix~\ref{sec:mmaudio_online}). Bold and underline mark the best and second-best scores within each group \red{(offline and online)}; \red{gray entries are quoted from the original paper (no official checkpoint) and are not ranked.} \red{The frontend look-ahead of WorldSonus (0.10\,s) equals three frames at 30\,fps.} \red{\textit{LiveFoley} uses History LocDiT as \red{the} default with $s=1$ \red{for both the 16\,kHz and 48\,kHz variants}.}}
\label{tab:main}
\begin{tabular*}{\textwidth}{@{\extracolsep{\fill}} l l c ccc cc cc @{}}
\toprule
\multirow{2}{*}{\textbf{Type}} & \multirow{2}{*}{\textbf{Method}} & \multirow{2}{*}{\shortstack{\textbf{Frontend}\\\textbf{Look-ahead (s)}}} & \multicolumn{3}{c}{\textbf{Distribution Matching}} & \multicolumn{2}{c}{\textbf{Audio Quality}} & \multicolumn{2}{c}{\textbf{Modality Alignment}} \\
\cmidrule(lr){4-6} \cmidrule(lr){7-8} \cmidrule(lr){9-10}
& & & $\mathrm{\textbf{FD}_{\textbf{PANNs}}}$↓ & $\mathrm{\textbf{FD}_{\textbf{PaSST}}}$↓ & $\mathrm{\textbf{KL}_{\textbf{PANNs}}}$↓ & $\mathrm{\textbf{IS}}$↑ & $\mathrm{\textbf{PQ}}$↑ & $\mathrm{\textbf{DeSync}}$↓ & $\mathrm{\textbf{IB}}$↑ \\
\midrule
\multirow{3}{*}{\textbf{Offline}}
& Frieren~\cite{frieren}   & 8.00 &\underline{12.13} & 123.75 & \underline{2.73} & 11.32 & 5.82 & \underline{0.86} & 0.21 \\
& AudioX~\cite{audiox}     & 10.00 &12.73 & \underline{121.82} & 2.96 & \textbf{13.34} & \underline{6.17} & 1.22 & \underline{0.24} \\
& MMAudio-L~\cite{mmaudio} & 8.00 &\textbf{8.48} & \textbf{74.16} & \textbf{2.18} & \underline{11.62} & \textbf{6.18} & \textbf{0.47} & \textbf{0.31} \\
\midrule
\multirow{6}{*}{\textbf{Online}}
& MMAudio (online)     & 0.64 & \red{\underline{9.82}} & 175.92 & \textbf{2.11} & 10.11 & 5.93 & 0.71 & \red{\underline{0.29}} \\
& V-AURA~\cite{vaura}      & 0.64 & 16.43 & 245.25 & \red{2.28} & 10.82 & 5.74 & 0.69 & 0.28 \\
& \textcolor{gray}{SoundReactor~\cite{soundreactor}} & \textcolor{gray}{0.08} & \textcolor{gray}{--} & \textcolor{gray}{113.00} & \textcolor{gray}{2.37} & \textcolor{gray}{--} & \textcolor{gray}{--} & \textcolor{gray}{1.03} & \textcolor{gray}{0.24} \\
& WorldSonus~\cite{worldsonus} & 0.10 & 16.32 & 197.68 & 2.54 & 8.73 & 6.11 & 0.67 & 0.26 \\
& \textbf{\textit{LiveFoley-16k (ours)}} & \red{0.08} & \red{\textbf{9.56}} & \red{\textbf{92.48}} & \red{\underline{2.24}} & \red{\underline{11.30}} & \red{\underline{6.14}} & \red{\underline{0.50}} & \red{0.29} \\
& \textbf{\textit{LiveFoley-48k (ours)}} & \red{0.08} & \red{9.92} & \red{\underline{128.34}} & \red{2.30} & \red{\textbf{11.42}} & \red{\textbf{6.26}} & \red{\textbf{0.48}} & \red{\textbf{0.29}} \\
\bottomrule
\end{tabular*}
\end{table*}

\paragraph{\red{Latency-Quality Trade-off.}}
\red{History LocDiT adds no \red{backend} look-ahead and serves as our default, while Future-Warmup LocDiT is an option for higher quality when latency permits. Even with its extra 0.08\,s backend look-ahead, the end-to-end latency at $s=1$ is about 236\,ms \red{(the measured 156\,ms plus the extra look-ahead; Appendix~\ref{sec:latency_appendix})}, still below the human perceptual threshold (Figure~\ref{fig:motivation}a).}

\subsection{Training and Inference}
\label{sec:training}

\paragraph{Training.}
The LM and LocDiT $u_\theta$ are jointly optimized via a conditional rectified-flow objective~\cite{rectifiedflow}:
\begin{equation}
\mathcal{L}(\theta) = \mathbb{E}\Big[\sum_{\red{k \in \mathcal{T}_i}} \big\| \red{u_\theta\big(x_{\mathcal{T}_i}^t; c_{\mathcal{T}_i}, x_{\mathcal{H}_i}^{0}\big)_k} - \big(\epsilon_k - x_k^{0}\big) \big\|_2^2\Big],
\label{eq:loss}
\end{equation}
where the expectation is over training data, autoregressive steps $i$, timesteps $t$, and noise $\epsilon$, and $x^t = (1-t)\,x^{0} + t\,\epsilon$ defines the interpolation trajectory ($t = 0$ for clean data, $t = 1$ for pure noise). \red{The target set $\mathcal{T}_i$ and the context set $\mathcal{H}_i$ of clean history patches are $\{i\}$ and $\{i{-}2,i{-}1\}$ for History LocDiT, and $\{i,i{+}1\}$ and $\{i{-}1\}$ for Future-Warmup LocDiT, whose $c_{i+1}$ is given in Eq.~\eqref{eq:future_warmup}; $(\cdot)_k$ denotes the output for patch $k$.} The timestep $t$ is sampled from a logit-normal distribution~\cite{sd3}, and \red{all target patches} in the same denoising window share a single $t$. To mitigate exposure bias during long-form generation, we apply noise injection~\cite{pasini2024cam,rouard2025calm}: the historical latent patches fed to LocEnc are perturbed with variance-preserving Gaussian noise during training; unlike \red{CALM}~\cite{rouard2025calm}, we also keep a small fixed noise level at inference (details in Appendix~\ref{sec:impl_appendix}).

\paragraph{Inference.}
To support streaming inference beyond the \red{9.92}\,s training sequences, we adopt an attention sink mechanism~\cite{streamingllm} with a rolling KV cache. The cache retains fixed initial tokens as attention sinks \red{that stabilize attention}, while a rolling window bounds memory usage. \red{Because KV entries are stored without RoPE~\cite{rope} and rotary embeddings are applied dynamically based on relative distance, sliding the window reduces to efficient pointer arithmetic and requires no recomputation of historical entries.}

\section{Experiments}
\label{sec:experiments}

\subsection{Experimental Setup}
\label{sec:setup}

\begin{figure*}[t]
\centering
\includegraphics[width=\linewidth]{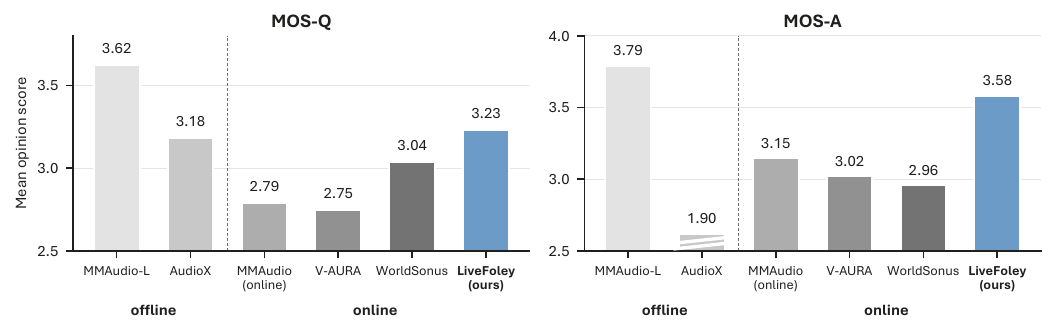}
\caption{Subjective evaluation of \textit{LiveFoley} against offline and online baselines: mean opinion scores for audio quality (MOS-Q, left) and audio-visual alignment (MOS-A, right). The dashed line separates offline from online systems. Protocol details are given in Appendix~\ref{sec:subjective_appendix}.}
\label{fig:subjective}
\end{figure*}

\paragraph{Training Data.}
Our training corpus contains about 2,000 hours (745K clips) of in-the-wild and egocentric video. In-the-wild clips come from SoundAtlas~\cite{omni2sound}, a VGGSound~\cite{vggsound} and AudioSet~\cite{audioset} corpus filtered for audio-visual consistency; egocentric clips come from EPIC-Sounds~\cite{epicsounds} and Ego4D-Sounds~\cite{ego4d,action2sound}, where we keep clips with high ImageBind~\cite{imagebind} audio-visual similarity. The full composition and preprocessing are given in Appendix~\ref{sec:data_appendix}. Our main results are evaluated on the standard VGGSound test set for comparability with prior work. However, VGGSound contains many visually unaligned sounds (\eg, off-screen narration and background music) that obscure the differences between visual frontends. We therefore construct VGGSound-Curated, a 1,485-clip evaluation subset with strict audio-visual alignment for visual frontend ablation experiments. Candidates are drawn from VGGSound-Sparse~\cite{sparsesync} and VisualSound~\cite{vaura} (ImageBind similarity above 0.3); we exclude clips containing speech or visible faces and keep only valid held-out test clips.

\paragraph{Evaluation Metrics.}
We evaluate on 8\,s clips with the av-benchmark toolkit~\cite{mmaudio} along four dimensions. Distribution matching is measured by the Fr\'echet distance between generated and ground-truth audio in the feature spaces of PANNs~\cite{panns} and PaSST~\cite{passt} ($\mathrm{FD}_{\mathrm{PANNs}}$ and $\mathrm{FD}_{\mathrm{PaSST}}$), and by the KL divergence between PANNs predictions ($\mathrm{KL}_{\mathrm{PANNs}}$). Audio quality is measured by the Inception Score (IS)~\cite{salimans2016improved} and the production quality score of Audiobox Aesthetics (PQ)~\cite{tjandra2025meta}. Semantic alignment between video and audio is measured by the ImageBind score (IB)~\cite{imagebind}, and temporal alignment by the desynchronization score predicted by Synchformer (DeSync)~\cite{synchformer}. For streaming systems, we further report the \red{frontend look-ahead} $\tau$, \ie, \red{the duration of video that must be buffered before the visual encoder can encode a new chunk or hop}, which is independent of hardware. For \textit{LiveFoley}, we also report the end-to-end latency, \ie, \red{the total wall-clock time from the arrival of the first video frame to the output of the first playable audio waveform}.

\paragraph{Implementation Details.}
The visual frontend is the frozen official Synchformer~\cite{synchformer}. We train two causal Oobleck VAEs~\cite{stableaudioopen}, a 16\,kHz mono one (25\,Hz) and a 48\,kHz stereo one (50\,Hz); with patch sizes of 2 and 4\red{, respectively}, the LM runs at 12.5\,Hz in both settings. The LM \red{follows the} 28-layer MiniCPM~\cite{minicpm,minicpm4} \red{architecture and is trained from scratch}\red{;} LocEnc and LocDiT are 12-layer transformers. \red{Table~\ref{tab:main} reports both variants; unless otherwise stated, the ablation studies use the 16\,kHz model with $s=2$.} Full architecture, training, and inference details are given in Appendix~\ref{sec:impl_appendix}.

\begin{table}[t]
\centering
\scriptsize
\setlength{\tabcolsep}{0pt}
\setlength{\aboverulesep}{1pt}
\setlength{\belowrulesep}{1pt}
\renewcommand{\arraystretch}{1.2}
\caption{Visual frontend comparison on VGGSound-Curated. Only the visual encoder changes\red{;} the V-AURA backbone, decoding, and training budget are identical\red{.} $\tau$ is the look-ahead of each encoder. Bold and underline mark the best and second-best scores.}
\vspace{-5pt}
\label{tab:encoder}
\begin{tabular*}{\linewidth}{@{\extracolsep{\fill}} c l c c c c c @{}}
\toprule
\textbf{Type} & \textbf{Visual encoder} & $\boldsymbol{\tau}$ (s) & $\mathrm{\textbf{FD}_{\textbf{PANNs}}}$↓ & $\mathrm{\textbf{FD}_{\textbf{PaSST}}}$↓ & $\mathrm{\textbf{KL}_{\textbf{PANNs}}}$↓ & $\mathrm{\textbf{DeSync}}$↓ \\
\midrule
\multirow{2}{*}{Frame}
& SigLIP2      & 0.08 & \textbf{27.92} & \underline{267.44} & \textbf{1.91} & 1.14 \\
& DINOv2       & 0.08 & 28.25 & 278.18 & 2.10 & 1.10 \\
\multirow{2}{*}{Segment}
& VideoMAE-V2  & 0.64 & 30.37 & 281.62 & 2.36 & \underline{0.88} \\
& \textbf{\textit{Synchformer (ours)}} & 0.08 & \underline{28.10} & \textbf{263.04} & \underline{2.01} & \textbf{0.61} \\
\bottomrule
\end{tabular*}
\end{table}

\subsection{Main Results}
\label{sec:main_results}
\paragraph{Quantitative Results.}
Table~\ref{tab:main} compares \textit{LiveFoley} with offline and streaming V2A systems on the full VGGSound test set. \textit{Alignment on par with offline models at a 0.08\,s \red{frontend} look-ahead:} \textit{LiveFoley}-48k reaches a DeSync of 0.48, on par with the offline MMAudio-L (0.47) and far ahead of all streaming baselines (0.67 to 1.03). \textit{Best audio quality among streaming systems:} \red{\textit{LiveFoley}-16k obtains the best $\mathrm{FD}_{\mathrm{PANNs}}$ (9.56) and $\mathrm{FD}_{\mathrm{PaSST}}$ (92.48) among streaming systems, and \textit{LiveFoley}-48k the best IS (11.42) among streaming systems and the highest PQ (6.26) of all systems}, and \red{the $\mathrm{FD}_{\mathrm{PaSST}}$ of \textit{LiveFoley}-16k} is second only to offline MMAudio-L (74.16). We attribute this \red{mainly} to the continuous-latent backbone and \red{the full-receptive-field visual condition, with History LocDiT further improving $\mathrm{FD}_{\mathrm{PANNs}}$} (Section~\ref{sec:ablation}). \textit{\red{A gap to offline models remains in generation quality:}} \textit{LiveFoley}-16k still trails MMAudio-L in \red{$\mathrm{FD}_{\mathrm{PaSST}}$ (92.48 vs.\ 74.16),} $\mathrm{FD}_{\mathrm{PANNs}}$ (\red{9.56} vs.\ 8.48), $\mathrm{KL}_{\mathrm{PANNs}}$ (\red{2.24} vs.\ 2.18), IS (\red{11.30} vs.\ 11.62), and IB (0.29 vs.\ 0.31). \red{Offline models jointly denoise all patches with global bidirectional attention, whereas a causal model must commit to each sound from the frames observed so far. Although Future-Warmup LocDiT further narrows this gap by bringing the upcoming patch into the current denoising window (Table~\ref{tab:backend}), closing it under the streaming constraint remains an open direction.}

\paragraph{Subjective Evaluation.}
We further conduct a listening test on offline and online systems (Figure~\ref{fig:subjective}; protocol in Appendix~\ref{sec:subjective_appendix}). The results largely agree with the objective metrics: \textit{LiveFoley} leads the online systems in both audio quality (MOS-Q 3.23) and audio-visual alignment (MOS-A 3.58), and approaches MMAudio-L in alignment (3.79). The exception is MMAudio (online), whose competitive $\mathrm{FD}_{\mathrm{PANNs}}$ and $\mathrm{KL}_{\mathrm{PANNs}}$ do not carry over to listening, as the seams of its training-free chunk-wise generation are clearly audible.

\subsection{Ablation Studies}
\label{sec:ablation}

\paragraph{Streaming Visual Frontend.}
In our experiments (Table~\ref{tab:encoder}), we compare different visual frontends for streaming V2A in terms of generation performance and their theoretical look-ahead. To isolate the visual frontend, we use V-AURA~\cite{vaura}, a widely adopted autoregressive V2A baseline, as the testbed on the VGGSound-Curated subset, keeping all other components identical. Results reveal a clear trade-off in existing methods. Frame-level encoders (\eg, DINOv2~\cite{dinov2}, SigLIP2~\cite{siglip2}) require almost no look-ahead but yield poor audio-visual alignment (DeSync $>$ 1.0), despite SigLIP2 providing strong semantics. Conversely, segment-level encoders model temporal context for better alignment; for instance, VideoMAE-V2~\cite{videomaev2} achieves a DeSync of 0.88 but incurs a high 0.64\,s look-ahead. \textit{Our proposed trailing sliding window breaks this trade-off, granting Synchformer segment-level alignment at frame-level latency.} With only 0.08\,s look-ahead, it attains a superior DeSync of 0.61 while \red{remaining competitive in distribution matching (best $\mathrm{FD}_{\mathrm{PaSST}}$ and second-best $\mathrm{FD}_{\mathrm{PANNs}}$ and $\mathrm{KL}_{\mathrm{PANNs}}$)}. Figure~\ref{fig:motivation}b further demonstrates \red{the robustness of our method} to latency reduction. For the chunked window, we train a separate Synchformer for each window size, and train the generation backend conditioned on the features of each model. As the look-ahead drops from 0.32\,s to 0.08\,s, DeSync degrades from \red{0.67} to \red{0.88} for the chunked window, whereas it remains nearly unchanged for the trailing sliding window (\red{0.60} to \red{0.61}).

\begin{table}[t]
\centering
\scriptsize
\setlength{\tabcolsep}{1.5mm}
\setlength{\aboverulesep}{1pt}
\setlength{\belowrulesep}{1pt}
\renewcommand{\arraystretch}{1.2}
\caption{Ablation of the Unified Varhop Framework on VGGSound-Curated. At each hop size $s$, our unified model is compared with a single-hop expert and with a unified variant trained without the Synchronizer. \red{$\tau$ is the frontend look-ahead.}}
\vspace{-5pt}
\label{tab:varhop}
\resizebox{\linewidth}{!}{%
\begin{tabular}{@{}l l cc c c c@{}}
\toprule
$\boldsymbol{s}$ ($\boldsymbol{\tau}$) & \textbf{Model} & $\mathrm{\textbf{FD}_{\textbf{PANNs}}}$↓ & $\mathrm{\textbf{FD}_{\textbf{PaSST}}}$↓ & $\mathrm{\textbf{KL}_{\textbf{PANNs}}}$↓ & $\mathrm{\textbf{DeSync}}$↓ & $\mathrm{\textbf{IB}}$↑ \\
\midrule
\multirow{3}{*}{1 (0.08)}
& Expert                & 17.07 & 140.22 & 1.78 & 0.47 & 0.33 \\
& \textbf{\textit{Unified (ours)}} & 16.94 & 145.02 & 1.90 & 0.47 & 0.32 \\
& \quad w/o Synchronizer & 17.31 & 147.76 & 1.88 & 0.50 & 0.32 \\
\midrule
\multirow{3}{*}{2 (0.16)}
& Expert                & 16.36 & 139.88 & 1.83 & 0.45 & 0.32 \\
& \textbf{\textit{Unified (ours)}} & 16.88 & 143.92 & 1.86 & 0.46 & 0.32 \\
& \quad w/o Synchronizer & 16.58 & 146.44 & 1.90 & 0.49 & 0.32 \\
\midrule
\multirow{3}{*}{4 (0.32)}
& Expert                & 17.04 & 143.38 & 1.83 & 0.45 & 0.32 \\
& \textbf{\textit{Unified (ours)}} & 17.11 & 143.33 & 1.82 & 0.45 & 0.32 \\
& \quad w/o Synchronizer & 16.79 & 146.15 & 1.87 & 0.47 & 0.31 \\
\bottomrule
\end{tabular}}
\vspace{-10pt}
\end{table}

\paragraph{Unified Varhop Framework.}
Table~\ref{tab:varhop} evaluates the Unified Varhop Framework across three predefined hop sizes against independent single-hop experts and a Synchronizer-free baseline. The unified model matches the experts in alignment \red{with only minor degradation in distribution metrics, so a single shared set of weights can support multiple latency modes and switch hop sizes dynamically during inference.} Furthermore, since the visual frontend runs once per window slide, setting $s=4$ reduces its computational overhead by $4\times$ compared to $s=1$. Finally, omitting the Synchronizer forces the model to naively slice the final $s$ tokens from the raw frontend output, which degrades both DeSync (by about 0.03) and $\mathrm{FD}_{\mathrm{PaSST}}$ at all hop sizes. This confirms the necessity of the Synchronizer: raw trailing tokens lack the intra-window position and context that its positional embedding and aggregation layers provide.

\paragraph{\red{Local Flow-Matching Head.}}
\red{Table~\ref{tab:backend} compares four denoising windows.
\textit{Widening toward the past helps:} History LocDiT \red{clearly improves $\mathrm{FD}_{\mathrm{PANNs}}$ (21.08$\to$17.55) and IB (\red{0.31}$\to$\red{0.32}) over Independent LocDiT, with comparable $\mathrm{KL}_{\mathrm{PANNs}}$, IS, and DeSync}.
\textit{Widening toward the future helps further, at an affordable cost:} both Low-Rate and Future-Warmup LocDiT access the upcoming patch and thus introduce an affordable backend look-ahead. With the same window length as History LocDiT, Future-Warmup LocDiT achieves the best distribution and quality scores (\eg, $\mathrm{FD}_{\mathrm{PaSST}}$ \red{158.04}$\to$\red{143.92}) and slightly improves DeSync (\red{0.47}$\to$\red{0.46}), consistent with the extra visual token it sees.
\textit{The full rate matters:} the halved temporal resolution of the backbone in Low-Rate LocDiT degrades alignment (DeSync \red{0.54}) and $\mathrm{FD}_{\mathrm{PaSST}}$ (\red{180.52}), both worse than Independent LocDiT.}

\begin{table}[t]
\centering
\scriptsize
\setlength{\tabcolsep}{0.8mm}
\setlength{\aboverulesep}{1pt}
\setlength{\belowrulesep}{1pt}
\renewcommand{\arraystretch}{1.2}
\caption{\red{LocDiT ablation} on VGGSound-Curated, comparing four denoising window designs under the same backbone. Bold and underline mark the best and second-best scores.}
\vspace{-5pt}
\label{tab:backend}
\resizebox{\linewidth}{!}{%
\begin{tabular}{l cc c c c c}
\toprule
\textbf{LocDiT window} & $\mathrm{\textbf{FD}_{\textbf{PANNs}}}$↓ & $\mathrm{\textbf{FD}_{\textbf{PaSST}}}$↓ & $\mathrm{\textbf{KL}_{\textbf{PANNs}}}$↓ & $\mathrm{\textbf{IS}}$↑ & $\mathrm{\textbf{DeSync}}$↓ & $\mathrm{\textbf{IB}}$↑ \\
\midrule
\red{Independent LocDiT}                        & 21.08 & \underline{154.62} & 1.94 & \underline{12.01} & 0.47 & 0.31 \\
Low-Rate LocDiT                & 18.90 & 180.52 & 2.01 & 10.38 & 0.54 & 0.30 \\
\red{History LocDiT}                   & \underline{17.55} & 158.04 & \underline{1.92} & 12.00 & \underline{0.47} & \textbf{0.32} \\
\red{Future-Warmup LocDiT} & \textbf{16.88} & \textbf{143.92} & \textbf{1.86} & \textbf{12.13} & \textbf{0.46} & \underline{0.32} \\
\bottomrule
\end{tabular}}
\vspace{-10pt}
\end{table}

\subsection{Deployment and Extensions}
\label{sec:deploy}
We further evaluate \textit{LiveFoley} along two axes. \textit{\red{End-to-end latency}:} \red{at $s=1$}, \textit{LiveFoley} achieves \red{156\,ms} on a single A800 GPU\red{, and Future-Warmup LocDiT adds 0.08\,s (236\,ms); both remain below the human perceptual threshold (Figure~\ref{fig:motivation}a; Appendix~\ref{sec:latency_appendix})}. \red{\textit{Multimodal control:} \textit{LiveFoley} extends to text prompts, with a single checkpoint \red{outperforming the state-of-the-art streaming system~\cite{worldsonus}} in both video-only and video-text modes (Appendix~\ref{sec:text_appendix}).}

\section{Conclusion}
\label{sec:conclusion}

We presented \textit{LiveFoley}, a streaming V2A framework that matches offline models in \red{temporal} alignment and narrows the gap in audio quality within human perceptual latency limits. The trailing sliding window turns non-causal segment-level encoders into streaming modules without retraining, and the Unified Varhop Framework lets a single model trade delay for computation. \red{In the backend, History LocDiT adds past patches to the denoising window at no latency cost, while the optional Future-Warmup LocDiT brings in the upcoming patch, further improving quality at a 0.08\,s backend look-ahead that the low-latency frontend makes affordable.} With a \red{156}\,ms \red{end-to-end latency} on a single A800 GPU, \textit{LiveFoley} can serve as a plug-and-play sound engine for interactive worlds.

\clearpage

{
    \small
    \bibliographystyle{ieeenat_fullname}
    \balance
    \ifdefined\nolinenumbers\nolinenumbers\fi
    \setlength{\bibsep}{0pt}
    \bibliography{main}
    \ifdefined\linenumbers\linenumbers\fi
}

\clearpage
\nobalance
\appendix
\section{User Study on Perceptual Delay Thresholds}
\label{sec:user_study}

\paragraph{Protocol.}
This study quantifies the perceptual delay threshold reported in Figure~\ref{fig:motivation}a, \ie, the minimum audio delay at which humans notice that the sound lags behind the video. Starting from clips with naturally synchronized sound and visuals, we progressively delay the audio stream by a controlled offset $D$ and identify the threshold at which participants begin to notice the mismatch. These thresholds serve as a conservative latency budget for streaming V2A, under the premise that an imperceptible delay does not disrupt user immersion. We examine three acoustic categories spanning a wide range of temporal sensitivity: action sounds synchronized with visible motion, speech, and ambient sound during free roaming. Nine clips (7--8\,s each) are selected across these categories: three action clips (tennis rally, wood chopping, and trumpet playing), three single-speaker talking-head clips, and three free-roaming scenes, sourced from the VGGSound~\cite{vggsound} test set, SpeakerVid-5M~\cite{speakervid5m}, and OGameData~\cite{gamegenx}\red{, respectively}.

Six participants \red{complete} the study with headphones, each adopting an active, first-person perspective (imagining performing the on-screen actions) and providing a binary response on whether the sound lagged behind the video; actual delay values \red{remain} hidden. The perceptual threshold is estimated via an independent adaptive staircase per clip, where the delay increases after an unperceived lag and decreases after a perceived one. Testing ranges are 0.1--0.5\,s for action, 0.2--0.8\,s for speech, and 0.4--1.0\,s for ambient scenes, with step sizes of 0.1--0.2\,s. Each staircase terminates after two directional reversals or a maximum of eight trials, and trials from different clips are interleaved randomly. A complete session involves 30--40 trials and lasts 7--15 minutes.

\paragraph{Metric and Results.}
For each clip and participant, the perceptual threshold is defined as the minimum delay at which the participant reports perceiving a lag. We report the mean and standard deviation across clips and participants within each category. Action sounds produce the lowest average threshold at $0.34 \pm 0.09$\,s (95\% CI over participant means $\pm 0.03$\,s), with the tennis rally at 0.27\,s (attributed to the sharp transients of ball impacts), wood chopping at 0.39\,s, and trumpet playing at 0.42\,s. Speech follows at $0.49 \pm 0.12$\,s ($\pm 0.10$\,s). Free-roaming scenes are tolerated up to $0.85 \pm 0.16$\,s ($\pm 0.10$\,s), though \red{in} 7 of 18 \red{clip-participant staircases the lag} \red{remains} undetected \red{even} at the 1.0\,s maximum, so this value is a lower bound on the true threshold. Paired $t$-tests confirm all pairwise category differences are significant ($p \leq 0.015$).

\red{\paragraph{Relation to Broadcast Standards.}
ITU-R BT.1359~\cite{itu_bt1359} reports a detectability threshold of about 125\,ms and an acceptability threshold of about 185\,ms for audio lagging video. These values are lower than ours, but the two studies measure different quantities and complement rather than contradict each other. BT.1359 asks viewers to compare a delayed sequence against an undelayed reference and rate the impairment of lip-synced speech, which yields a strict bound for broadcast quality control. In interactive applications, however, no synchronized reference is available, and users judge synchrony from what they see and hear while engaged with the content. Our protocol follows this setting: participants judge each clip without a reference and report whether they notice a lag, and the content covers action sounds, speech, and ambient scenes. As a reference makes small offsets easier to spot, BT.1359 serves as a stricter bound, whereas our results reflect when a lag becomes noticeable in natural viewing. The default end-to-end latency of 156\,ms is acceptable under both views: it is below our measured thresholds and within the acceptability threshold of BT.1359.}

\section{Dataset Details}
\label{sec:data_appendix}

Table~\ref{tab:train_data} lists the composition of our training set. SoundAtlas~\cite{omni2sound} contributes the VGGSound and AudioSet~\cite{audioset} portions, both filtered for audio-visual consistency. EPIC-Sounds~\cite{epicsounds} and Ego4D-Sounds~\cite{action2sound} provide egocentric footage; from each we keep clips with high audio-visual correspondence measured by ImageBind~\cite{imagebind} similarity. All videos are decoded at 25\,fps and trimmed to 10\,s, and clips are further cropped to at most 9.92\,s (124 patches at 12.5\,Hz) during training. At evaluation, all systems receive the same 25\,fps videos, except WorldSonus~\cite{worldsonus}, which is run at 30\,fps following its official setup.

\begin{table}[t]
\centering
\footnotesize
\setlength{\tabcolsep}{3.4mm}
\setlength{\aboverulesep}{1pt}
\setlength{\belowrulesep}{1pt}
\renewcommand{\arraystretch}{1.15}
\caption{Training data composition.}
\label{tab:train_data}
\begin{tabular}{l l r r}
\toprule
\textbf{Source} & \textbf{Domain} & \textbf{Clips} & \textbf{Hours} \\
\midrule
SoundAtlas (VGGSound)  & in-the-wild & 146K & 401 \\
SoundAtlas (AudioSet)  & in-the-wild & 340K & 929 \\
EPIC-Sounds            & egocentric  & 49K  & 134 \\
Ego4D-Sounds           & egocentric  & 210K & 578 \\
\midrule
\textbf{Total}         &             & \textbf{745K} & \textbf{2,042} \\
\bottomrule
\end{tabular}
\end{table}

\section{Implementation Details}
\label{sec:impl_appendix}

\begin{figure*}[t]
  \centering
  \includegraphics[width=0.92\textwidth]{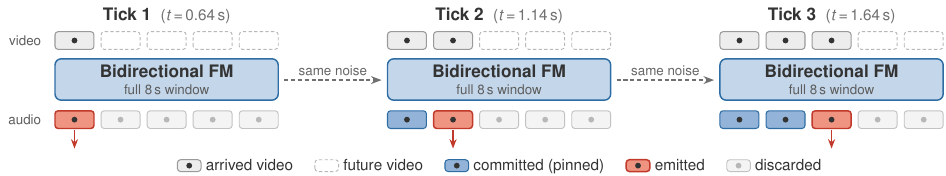}
  \caption{Training-free online adaptation of MMAudio (first three ticks shown). At each tick, the pretrained bidirectional flow-matching (FM) model is rerun over its full 8\,s window, conditioned only on the video that has arrived. Committed audio chunks are pinned during denoising, the next chunk is emitted after the run, and the remaining chunks are discarded and regenerated at the next tick. Dashed arrows indicate that the same initial noise is carried across ticks.}
  \label{fig:mmaudio_online}
\end{figure*}

\paragraph{Architecture.} The visual frontend is the official Synchformer~\cite{synchformer} checkpoint, kept frozen during training. Both audio VAEs adopt the Oobleck architecture of Stable Audio Open~\cite{stableaudioopen}: the 16\,kHz mono VAE uses a temporal downsampling factor of 640 (25\,Hz, 64-d latents), and the 48\,kHz stereo VAE uses a factor of 960 with a causal transformer bottleneck (50\,Hz, 64-d latents). Each VAE is first pretrained with the standard non-causal recipe and then fine-tuned after replacing all convolutions with causal ones to avoid look-ahead latency. With patch sizes of 2 and 4 latent frames, respectively, the LM runs at 12.5\,Hz in both settings. The LM \red{follows the} 28-layer MiniCPM~\cite{minicpm,minicpm4} \red{architecture} with a hidden size of 2048 \red{and is trained from scratch}; LocEnc and LocDiT are 12-layer transformers with a hidden size of 1024 and 16 attention heads. At inference, the streaming engine runs a three-stage asynchronous pipeline (frontend, LM with LocDiT, and VAE decoding) on separate threads.

\paragraph{Training and Inference.} We train with AdamW~\cite{adamw} ($\beta=(0.9,0.99)$, weight decay 0.05)\red{,} a constant learning rate of $10^{-4}$ after 1K warmup steps, gradient clipping \red{at} 1.0, and a batch \red{size} of 64 clips on \red{16 A800} GPUs for 200K steps. Classifier-free guidance~\cite{cfg} is trained following DiTAR~\cite{ditar} with a condition dropout of 0.1. As data augmentation, the visual condition of LocDiT is dropped with probability 0.1\red{. For} the noise injection \red{in} Section~\ref{sec:training}, each historical latent patch fed to LocEnc is \red{independently replaced, with probability 0.5, by} a variance-preserving mixture $\tilde{x} = \sqrt{\gamma}\,\eta + \sqrt{1-\gamma}\,x^{0}$, where $\gamma \sim \mathcal{U}(0, 0.5)$ and $\eta \sim \mathcal{N}(0, I)$\red{;} the \red{history patches} and targets of LocDiT remain clean. At inference, each generated patch is mixed with a fixed $\gamma = 0.15$ before entering LocEnc, which we find \red{improves} generation quality\red{, and we} use a guidance scale of 6 and 20 Euler denoising steps. For long-form streaming, the LM keeps a rolling KV cache of 124 tokens, matching the 9.92\,s training length at 12.5\,Hz, plus 4 attention-sink tokens pinned at the beginning of the stream~\cite{streamingllm}. Keys are cached without rotary embeddings, and RoPE~\cite{rope} is applied at attention time from the relative distance between query and key, so evicting the oldest entries requires no recomputation of the remaining cache.

\section{Online Adaptation of MMAudio}
\label{sec:mmaudio_online}

MMAudio~\cite{mmaudio} is an offline flow-matching model that denoises an entire 8\,s window at once with bidirectional attention, conditioned on the CLIP~\cite{radford2021clip} and Synchformer features of the whole clip. We convert the pretrained MMAudio-L into a streaming baseline without retraining, keeping the 8\,s window and the Euler sampler unchanged; Figure~\ref{fig:mmaudio_online} illustrates the procedure.

\paragraph{Causal Conditioning.}
The video is consumed tick by tick, and at each tick only the frames that have arrived are encoded. CLIP features are extracted per frame and Synchformer features per 16-frame (0.64\,s) segment; both are extracted incrementally, so each tick encodes only the newly arrived frames and segments. The conditioning sequence retains the length expected by the model: positions corresponding to future frames are filled by repeating the last available token, so no future information leaks in. Because the first Synchformer segment is complete only after 16 frames, the first tick waits 0.64\,s, which is the \red{frontend look-ahead} reported in Table~\ref{tab:main}.

\paragraph{Committed Prefix with Pinned Denoising.}
The window does not slide. Generation proceeds in ticks: the first tick starts once 0.64\,s of video has arrived, and each subsequent tick once a further 0.5\,s has arrived. At every tick, the sampler runs the full Euler trajectory over the entire 8\,s window, conditioned on the video observed so far and starting from the same fixed initial noise. Latents committed in earlier ticks are pinned in the style of RePaint~\cite{repaint}: at each Euler step they are reset to the forward-noised version of the committed audio at the current timestep, and at the final step to the committed audio itself. After the run, only the latents corresponding to the newly arrived video are committed and decoded, whereas the latents beyond them, which were denoised under repeat-padded conditions, are discarded and regenerated at the next tick. This process repeats until the whole window is committed. Reusing the same initial noise across ticks allows each tick to continue the same underlying sample, which reduces discontinuities at chunk boundaries, although residual seams remain audible in listening tests (Section~\ref{sec:main_results}).

\section{\red{End-to-End Latency Breakdown}}
\label{sec:latency_appendix}

\begin{table}[t]
\centering
\footnotesize
\setlength{\tabcolsep}{9.5mm}
\setlength{\aboverulesep}{1pt}
\setlength{\belowrulesep}{1pt}
\renewcommand{\arraystretch}{1.15}
\caption{\red{End-to-end latency breakdown} on a single A800 GPU (\red{default setting: History LocDiT, hop $s=1$}). \red{The end-to-end latency} is measured with a real-time simulator that releases video frames at 25\,fps.}
\label{tab:latency}
\begin{tabular}{l r}
\toprule
\textbf{Stage} & \textbf{Time} \\
\midrule
\red{Frontend look-ahead} ($\tau$, \red{$s=1$})   &\red{80}\,ms \\
Visual frontend                 & 18\,ms  \\
LM step                         & 17\,ms  \\
LocDiT denoise                  & 25\,ms  \\
Causal VAE decode               & 8\,ms   \\
Pipeline overhead               & 8\,ms   \\
\midrule
\textbf{\red{End-to-end latency}} & \textbf{\red{156}\,ms} \\
\bottomrule
\end{tabular}
\par\vspace{10pt}
\includegraphics[width=\linewidth]{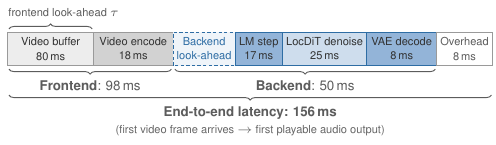}
\captionof{figure}{\red{Composition of the end-to-end latency in the default setting, from the arrival of the first video frame to the output of the first playable audio. The frontend buffers and encodes one hop of video, and the backend runs one LM step, LocDiT denoising, and VAE decoding. The video buffer is the frontend look-ahead $\tau$; the backend look-ahead (dashed) is zero for History LocDiT and 80\,ms for Future-Warmup LocDiT. Block widths are not to scale.}}
\label{fig:latency_timeline}
\end{table}

Table~\ref{tab:latency} \red{and Figure~\ref{fig:latency_timeline} decompose} the \red{end-to-end latency} of \textit{LiveFoley} on an A800 GPU \red{in the default setting}. \red{The frontend look-ahead $\tau$} is an algorithmic wait that does not depend on hardware; \red{the remaining compute stages} together take under 80\,ms. \red{Table~\ref{tab:lookahead} extrapolates the latency for every hop size and both LocDiT variants from this measurement.}

\begin{table}[t]
\centering
\footnotesize
\setlength{\aboverulesep}{1pt}
\setlength{\belowrulesep}{1pt}
\renewcommand{\arraystretch}{1.6}
\caption{\red{Frontend look-ahead, backend look-ahead, and end-to-end latency for each hop size $s$, reported as frontend look-ahead (s) / backend look-ahead (s) / end-to-end latency (ms). Only History LocDiT at $s=1$ (156\,ms, bold) is measured; the other latencies are obtained by adding the extra algorithmic look-ahead to this measurement, assuming the same per-patch compute.}}
\label{tab:lookahead}
\begin{tabular*}{\linewidth}{@{\extracolsep{\fill}} l ccc @{}}
\toprule
\textbf{LocDiT} & $\boldsymbol{s=1}$ & $\boldsymbol{s=2}$ & $\boldsymbol{s=4}$ \\
\midrule
History       & 0.08/0.00/\textbf{156} & 0.16/0.00/236 & 0.32/0.00/396 \\
Future-Warmup & 0.08/0.08/236          & 0.16/0.08/316 & 0.32/0.08/476 \\
\bottomrule
\end{tabular*}
\end{table}

\section{Text Conditioning Extension}
\label{sec:text_appendix}

The \textit{LiveFoley} variant presented in the main paper operates purely on video input. This video-only design reflects our target scenarios, which are dynamic, interactive environments such as 3D game engines and pre-rendered virtual worlds, where audio must be generated at the pace of the rendered frames and no latency budget is available for an external captioner (\eg, a VLM) to produce a text prompt at inference time. In other content-generation settings, such as generation-based world models, however, a text prompt is typically available in advance and can provide useful semantic guidance at no latency cost. In this section, we therefore extend \textit{LiveFoley} with text as an optional conditioning modality, so that a single checkpoint supports both a video-only mode and a joint video-and-text mode.

\paragraph{Setup.} We train a text-conditioned variant of \textit{LiveFoley}. Text supervision is extracted from the per-clip captions provided by SoundAtlas~\cite{omni2sound}. Text embeddings are produced by a frozen Flan-T5~\cite{flant5} encoder and injected into the generative backbone through cross-attention, similar to WorldSonus~\cite{worldsonus}. Training mixes text-conditioned and video-only samples, so a single checkpoint serves both modes. During inference, each video clip is accompanied by its corresponding text prompt.

\paragraph{Results.} As shown in Table~\ref{tab:text}, \textit{LiveFoley} maintains a clear lead over the concurrent WorldSonus in both the video-only and the joint video-and-text modes, on all metrics except $\mathrm{KL}_{\mathrm{PANNs}}$ with text (1.58 \red{and 1.61} vs.\ 1.51), confirming that \textit{LiveFoley} extends naturally to text as an optional conditioning modality. For both models, the semantic guidance from text mainly improves distribution matching (for \textit{LiveFoley}, $\mathrm{FD}_{\mathrm{PANNs}}$ 9.61$\to$7.30 and $\mathrm{KL}_{\mathrm{PANNs}}$ 2.47$\to$1.58) at a mild cost in audio-visual alignment (DeSync 0.47$\to$0.50). Informal listening corroborates this trade-off: text yields more stable and semantically coherent audio, while temporal alignment is slightly weakened.

\begin{table*}[t]
\centering
\footnotesize
\setlength{\tabcolsep}{0pt}
\setlength{\aboverulesep}{1pt}
\setlength{\belowrulesep}{1pt}
\caption{Text-conditioning comparison on the VGGSound test set. \red{For each of \textit{LiveFoley}-16k and \textit{LiveFoley}-48k,} both rows come from the same checkpoint, trained on a mixture of text-conditioned and video-only samples and evaluated without and with per-clip text prompts. \red{Their} w/o-text scores therefore differ from \red{those in} Table~\ref{tab:main}, whose \textit{LiveFoley} model is trained on video only. All other \red{evaluation} settings follow Table~\ref{tab:main}; bold marks the best score in each column.}
\label{tab:text}
\begin{tabular*}{\textwidth}{@{\extracolsep{\fill}} l ccc cc cc @{}}
\toprule
\textbf{Method} & $\mathrm{\textbf{FD}_{\textbf{PANNs}}}$↓ & $\mathrm{\textbf{FD}_{\textbf{PaSST}}}$↓ & $\mathrm{\textbf{KL}_{\textbf{PANNs}}}$↓ & $\mathrm{\textbf{IS}}$↑ & $\mathrm{\textbf{PQ}}$↑ & $\mathrm{\textbf{DeSync}}$↓ & $\mathrm{\textbf{IB}}$↑ \\
\midrule
WorldSonus~\cite{worldsonus} w/o text  & 16.32 & 197.68 & 2.54 & 8.73  & 6.11 & 0.67 & 0.26 \\
WorldSonus~\cite{worldsonus} w/ text   & 9.96  & 193.75 & \textbf{1.51} & 9.89  & 6.19 & 0.76 & 0.26 \\
\textit{LiveFoley\red{-16k}} w/o text (ours)     & 9.61  & 90.54  & 2.47 & 11.36 & \textbf{6.44} & \textbf{0.47} & 0.29 \\
\textit{LiveFoley\red{-16k}} w/ text (ours)      & \textbf{7.30} & \textbf{82.51} & 1.58 & \textbf{11.88} & 6.36 & 0.50 & \textbf{0.30} \\
\red{\textit{LiveFoley-48k} w/o text (ours)} & \red{10.16} & \red{129.48} & \red{2.45} & \red{10.95} & \red{6.39} & \red{0.48} & \red{0.28} \\
\red{\textit{LiveFoley-48k} w/ text (ours)}  & \red{7.34}  & \red{122.86} & \red{1.61} & \red{11.21} & \red{6.38} & \red{0.49} & \red{0.29} \\
\bottomrule
\end{tabular*}
\end{table*}

\section{Subjective Evaluation}
\label{sec:subjective_appendix}

\paragraph{Protocol.}
To complement the objective metrics in Table~\ref{tab:main}, we conduct a listening test comparing \textit{LiveFoley} against two offline systems (MMAudio-L and AudioX~\cite{audiox}) and three online systems (MMAudio (online), V-AURA~\cite{vaura}, and WorldSonus~\cite{worldsonus}). SoundReactor~\cite{soundreactor} is excluded because no official checkpoint is available. We select 16 clips from two categories: eight in-the-wild clips from the VGGSound test set and eight game-engine clips from YouTube covering both fast-paced action and ambient open-world traversal. Each clip is trimmed to 10\,s with the original audio removed, so all systems process identical visual inputs in a video-only setting\red{, each in its streaming configuration. WorldSonus uses its official checkpoint without a text prompt, and} V-AURA runs in streaming mode with a rolling KV cache and attention sinks, matching the long-form streaming protocol described in Appendix~\ref{sec:impl_appendix}. The generated audio is recombined with the source video, truncated to 8\,s, and loudness-normalized to $-23$\,LUFS via linear gain to prevent volume differences from biasing the evaluation. Six participants rate the outputs on desktop browsers with headphones. For each clip, all six systems are displayed side-by-side in anonymized, randomly shuffled order (Model~A through~F). Participants may freely replay and compare before scoring each output on a 5-point Likert scale (1: poor/misaligned; 5: excellent/perfectly aligned) for the two criteria defined below.

\paragraph{Metrics.}
\textbf{MOS-Q (audio quality).} This metric evaluates intrinsic audio quality independent of the video. Raters \red{assess} naturalness, clarity, mixing comfort, and the absence of artifacts such as noise or distortion. \textbf{MOS-A (audio-visual alignment).} This metric assesses how well the sound matches the visual content, encompassing semantic and temporal alignment. Semantically, acoustic sources must correspond to on-screen elements (\eg, rain sounding like rain rather than traffic). Temporally, sound events must synchronize precisely with visual actions, such as a footstep aligning exactly with a foot strike, without noticeable \red{delay or early onset}. In summary, MOS-Q measures standalone audio fidelity, whereas MOS-A evaluates coherence between the audio and visual modalities.

\paragraph{Results.}
Figure~\ref{fig:subjective} in the main paper summarizes the subjective evaluation results. Among online systems, \textit{LiveFoley} consistently \red{outperforms} the baselines across both criteria, achieving a MOS-Q of 3.23 (compared to 2.79 for MMAudio (online), 2.75 for V-AURA, and 3.04 for WorldSonus) and a MOS-A of 3.58 (surpassing 3.15, 3.02, and 2.96, respectively). Notably, WorldSonus operates at a comparable \red{frontend look-ahead} (0.10\,s) yet trails \textit{LiveFoley} most clearly in alignment. Compared with offline models, \textit{LiveFoley} approaches MMAudio-L in audio-visual alignment (3.58 versus 3.79), though a gap in audio quality remains (3.23 versus 3.62). While AudioX \red{achieves} an audio quality comparable to \textit{LiveFoley}, it \red{receives} the lowest alignment score (1.90), corroborating \red{its high DeSync (1.22)} in Table~\ref{tab:main}. Overall, the subjective rankings are consistent with the objective metrics: \textit{LiveFoley} closes much of the alignment gap with offline models while operating at a \red{frontend look-ahead} \red{well below the perceptual threshold of action sounds}.

\end{document}